\documentclass{osa-article}

\journal{osajournal}

\articletype{Research Article}

\begin{document}
\title{Spatial power spectrum of natural water turbulence with any average temperature, salinity concentration and light wavelength}

\author{Jin-Ren Yao,\authormark{1} Mohammed Elamassie,\authormark{2} Olga Korotkova,\authormark{3,*}}

\address{\authormark{1}School of Physics, Harbin Institute of technology, Harbin, 150001, China\\
\authormark{2}Department of Electrical and Electronics Engineering, \"{O}zye\v{g}in University, 34794 Istanbul, Turkey\\
	\authormark{3}Department of Physics, University of Miami, Coral Gables, FL 33146, USA}

\email{\authormark{*}Corresponding author: korotkova@physics.miami.edu} 

\begin{abstract}
The power spectrum of water optical turbulence is shown to vary with its average temperature $\left\langle T \right\rangle$ and average salinity concentration $\left\langle S \right\rangle$, as well as with light wavelength $\lambda$. This study explores such variations for $\left\langle T \right\rangle  \in \left[ {0\ ^\circ\rm{C},30\ ^\circ\rm{C}} \right]$, $\left\langle S \right\rangle  \in \left[ {0\ \rm{ppt},40\ \rm{ppt}} \right]$ covering most of the possible natural water conditions within the Earth's boundary layer and for visible electromagnetic spectrum, $\lambda \in \left[400\ nm, 700\ nm \right]$. For illustration of the effects of these parameters on propagating light we apply the developed power spectrum model for estimation of the scintillation index of a plane wave (the Rytov variance) and the threshold between weak and strong turbulence regimes.
\end{abstract}

\section{Introduction}
The demand for underwater applications has been steadily increasing in the recent years due to broad expansion in human activities such as scientific data collection, environmental monitoring, oil field exploration, maritime archaeology, port and watercraft security. In their turn, these applications have boosted the demand for the underwater high speed wireless connectivity and high quality imaging. Optical signaling has the ability to achieve these goals but it is severely affected by water optical turbulence, i.e., rapid but spatially mild fluctuations in the water's refractive index \cite{KorOTreview} (see also \cite{Thorpe, KorRB}). Therefore it is crucial to establish accurate analytical models for spatial power spectrum applicable to the wide range of Earth's water conditions. It is well known that the two main factors affecting water optical turbulence are the temperature and the salinity fluctuations. However, unlike in air turbulence, in the water not only the variations in these quantities but also their average values may affect the optical signal transmission. Moreover, unlike in the air, on propagation in water light statistics may substantially depend on the source wavelength.

Originally the spatial power spectra models of the fluctuating refractive index of the oceanic waters resulting from temperature and salinity fluctuations have been developed in the seminal work by Hill in 1978 \cite{Hill1}, \cite{Hill2}. Not until 2000 have the two power spectra been combined into a single power spectrum in a form of a linearized polynomial by Nikishovs \cite{Nikishovs}, involving some data from the previously developed model for double diffusers \cite{Ruddick}. Due to its simplicity and versatility, over the last two decades the Nikishovs' model became the standard for making the theoretical predictions about light evolution in underwater turbulence \cite{KorOTreview}. Notwithstanding its long-lasting impact on the field, several elaborations of the Nikishovs' spectrum have been later proposed \cite{Elamassie,Yao,Yi,YaoKor}. 

In particular, in recent work by the authors \cite{YaoKor} (see also \cite{YaoKorLIDAR}) the attempt was made to use a numerical fit to one of the very accurate Hill's models (model 4 of \cite{Hill1}) for the power spectrum with Prandtl(Pr)/Schmidt(Sc) numbers varying in intervals sufficiently large to cover all possibile average temperatures occuring in the Earth's ocean waters: $\left<T\right> \in [0\ ^\circ \rm{C}, 30\ ^\circ \rm{C}]$. The effect of the average temperature on the evolution of the light waves was then revealed: with the increase in the average temperature the light statistics have been shown to be affected slightly less. The proposed extension \cite{YaoKor}, while provided the insight into the average temperature dependence has only dealt with the ocean waters at the average salinity (NaCl) concentration of $35\rm{ppt}$. 

This study, building upon model in \cite{YaoKor}, considers turbulent waters with large ranges of average temperature $\left<T\right>\in[0\ ^\circ \rm{C}, 30\ ^\circ \rm{C}]$ and average salinity concentration $\left<S\right>\in[0\ \rm{ppt}, 40\ \rm{ppt}]$ covering practically all possible water basins on the planet. In particular, here we introduce two important extensions: (I) we calculate the Prandtl/Schmidt numbers and the eddy diffusivity ratio for arbitrary average temperature and salinity concentration and (II) we recalculate the linear coefficients used in the Nikishovs' linearized polynomial model, on the basis of a precise model of the refractive index of the ocean water developed by Quan and Fry \cite{QuanFry}. The latter extension allows us to determine these linear coefficients having fine dependence on the average temperature, average salinity concentration and the wavelength of light. We point out that the direct dependence of the underwater power spectrum on the light wavelength is a new type of dependence and will be examined in detail in terms of its impact on light scintillation. In fact, this should not come as a surprise: water absorption has a very strong dependence on light wavelength as well \cite{PopeFry}.         

\section{Calculation of temperature/salinity dependent Prandtl and Schmidt numbers} 

In this section, we analyze the dependence of the Prandtl and the Schmidt numbers on the water's average temperature and salinity concentration. In the Hill's models  \cite{Hill1,Hill2} these parameters are calculated on the basis of the laminar flows and must not be confused with similar definitions for turbulent flows. We will assume here that the calculations of all derived quantities are made at the atmospheric pressure. 

The \textit{Prandtl number} for the laminar flow is defined as
\begin{equation}
	Pr=\nu / \alpha_T, 
	\label{eq1}
\end{equation}
where $\nu \left[\rm{m^{2}s^{-1}}\right]$ is the momentum diffusivity (also known as kinematic viscosity), $\alpha_T$ [m$^2$/s] is the molecular thermal diffusivity. Further, $\nu$ can be expressed as
\begin{equation}
	\nu=\mu/\rho,
	\label{eq2}
\end{equation}
with $\mu$ being dynamic viscosity $\left[\rm{N\cdot s\cdot m^{2}}\right]$  and $\rho \left[\rm{kg\cdot m^3}\right]$ being the density of a fluid. Also in Eq. \eqref{eq1} $\alpha_T$ is
\begin{equation}
	\alpha_T=\sigma_T/(\rho c_p),
	\label{eq3}
\end{equation}
with $\sigma_T\left[\rm{W\cdot m^{-1}K^{-1}}\right]$ being the thermal conductivity and $c_p \left[\rm{J\cdot kg^{-1}\cdot K^{-1}}\right]$ being the specific heat. Hence, 
\begin{equation}
	Pr=c_p \mu/\sigma_T.
	\label{eq4}
\end{equation}
As shown in Appendices I-III, three parameters: $c_p$, $\mu$, $\sigma_T$ may be directly related to the average temperature $\langle T\rangle$ and average salinity $\langle S\rangle$. 
\textit{Thus, on combining Eq. (\ref{eq4}) with Eqs. (\ref{eq29}-\ref{eq35}), we have directly related $Pr$ with $\left<T\right>$ and $\left<S\right>$.}

The \textit{Schmidt number} for the laminar flow is defined as
\begin{equation}
	Sc=\nu/\alpha_S,
	\label{eq5}
\end{equation}
where, as before, $\nu$ is kinematic viscosity and $\alpha_S\left[\rm{m^{2}s^{-1}}\right]$ is the molecular diffusivity of salt. 
According to the Stokes--Einstein law \cite{poisson1983},
	\begin{equation}
	{{\alpha _S}\mu }/\langle T \rangle = {\rm{constant}}.
	\label{eq6}
	\end{equation}
On fitting the data in Ref. \cite{poisson1983} with the least square method we concluded that
	\begin{equation}
	{\alpha _S} \approx {\rm{5}}.{\rm{954}} \times {10^{ - 15}}\frac{\langle T \rangle}{\mu }.
	\label{eq7}
	\end{equation}
Further, on combining Eq. (\ref{eq7}) with Eqs. (\ref{eq2}) and (\ref{eq5}) we arrive at the formula
\begin{equation}
	Sc = \frac{\mu }{{{\alpha _S}\rho }} \approx \frac{{{\mu ^2}}}{{{\rm{5}}.{\rm{954}} \times {{10}^{ - 15}}\left\langle T \right\rangle \rho }}.
	\label{eq8}
\end{equation}
The details regarding variation of $\mu$ and $\rho$ with $\langle T\rangle$ and $\langle S\rangle$ are given in Appendices III and IV, respectively. \textit{Thus, on combining Eq. (\ref{eq8}) with Eqs. (\ref{eq33})-(\ref{eq38}) we have directly related $Sc$ with $\left<T\right>$ and $\left<S\right>$}.

\begin{figure}
	\centering
	\includegraphics[width=0.4\textwidth]{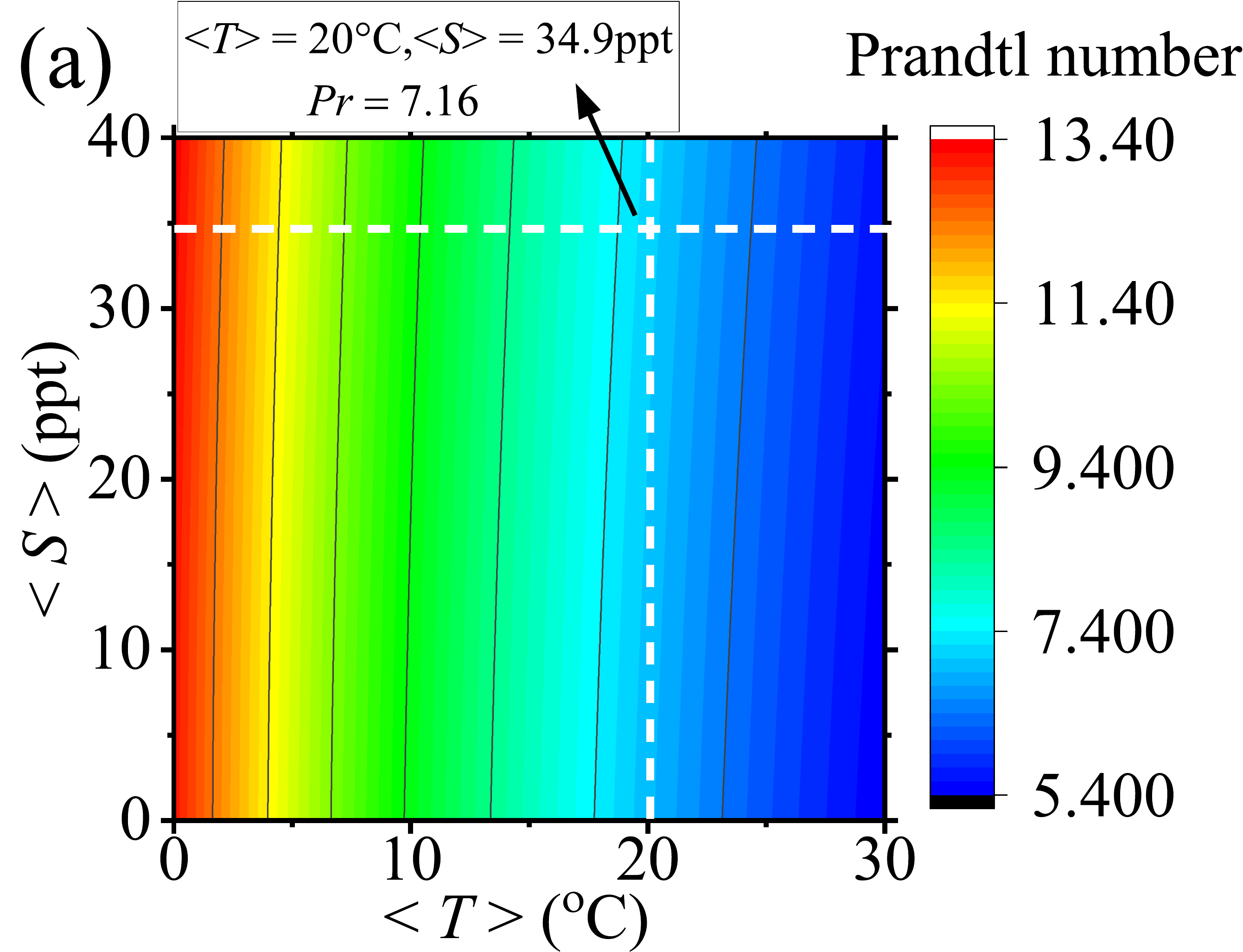}
	\includegraphics[width=0.4\textwidth]{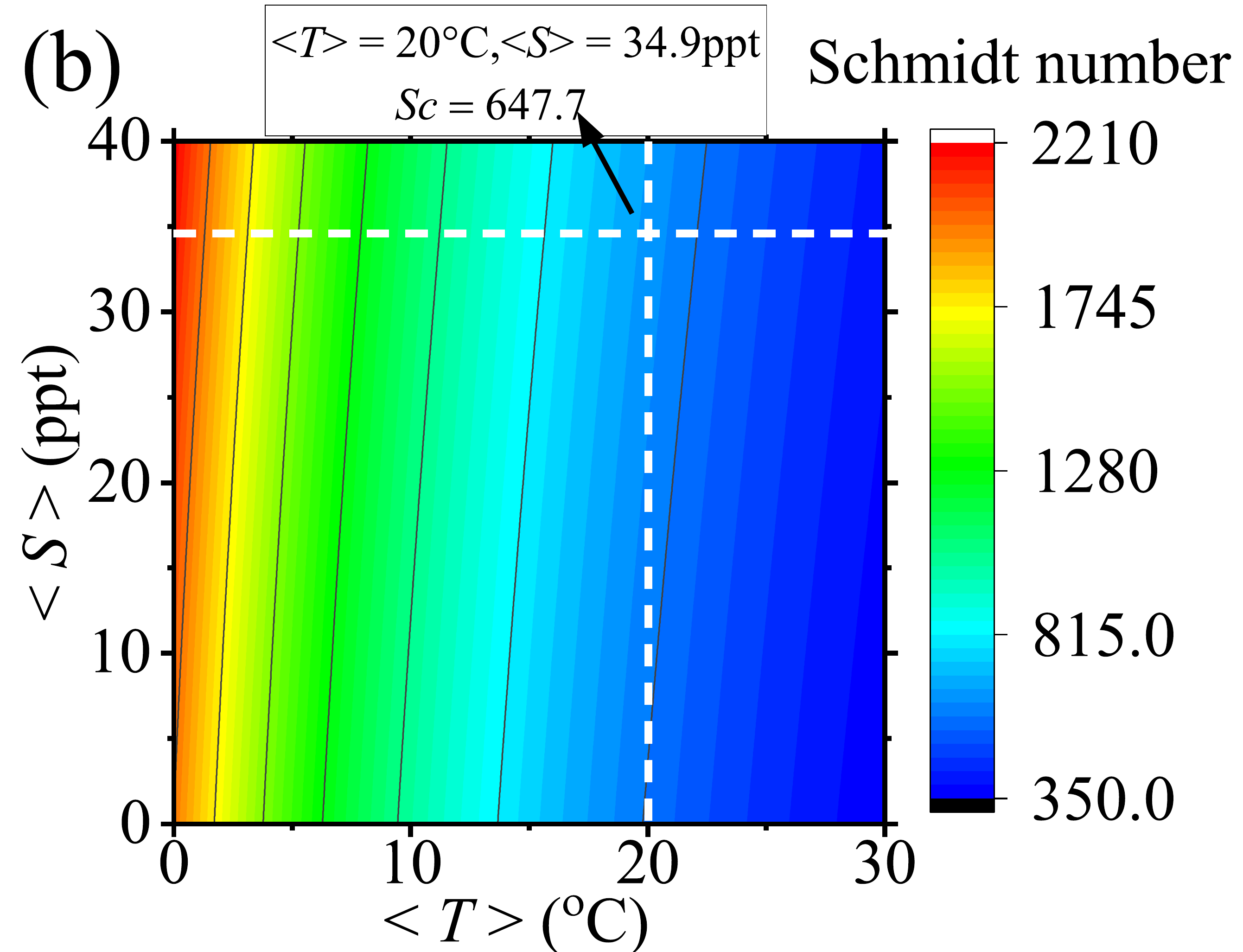}
	\caption{(a) Schmidt number, and (b) Prandtl number varying with average temperature and salinity.}
	\label{fig:PrSc}
\end{figure}

Figure \ref{fig:PrSc} uses the results of this section by presenting the density plots of the Schmidt and the Prandtl numbers varying with the average temperature and salinity concentration. 
The state $(\left<T\right>,\left<S\right>)=(20^\circ\rm{C},34.9\rm{ppt})$ corresponding to  $Pr=7.16$ and $Sc=647.7$ sufficiently agrees with the widely accepted approximation of standard oceanic turbulence ($Pr\approx7$ and $Sc\approx700$). Both $Pr$ and $Sc$ decrease with increasing $\langle T\rangle$, and slowly increase with increasing $\langle S\rangle$ being in agreement with that of Ref. \cite{YaoKor,YaoKorLIDAR}. 

\section{Calculation of temperature/salinity dependent eddy diffusivity ratio} 

The aim of this section is to obtain the expression for the eddy diffusivity ratio $d_r$ varying with water's average temperature and average salinity \cite{footnote1}. This quantity is defined as  \cite{Elamassie,YaoKor}
\begin{equation}
{d_r}{{ = {K_S}} \mathord{\left/
		{\vphantom {{ = {K_S}} {{K_T}}}} \right.
		\kern-\nulldelimiterspace} {{K_T}}} \approx \left\{ {\begin{array}{*{20}{l}}
	{R_\rho + R_\rho^{0.5}{{(R_\rho - 1)}^{0.5}},}&{R_\rho \ge 1,}\\
	{1.85 R_\rho - 0.85,}&{0.5 \le R_\rho < 1,}\\
	{0.15 R_\rho,}&{R_\rho < 0.5,}
	\end{array}} \right.
\label{eq9}
\end{equation}
where $K_T$ and $K_S$ are the eddy diffusivity of temperature and salinity, respectively. Further,  $R_\rho$ is a dimensionless quantity known as the density ratio,
\begin{equation}
{R_\rho }  = \frac{\alpha \left| H \right| }{\beta },
\label{eq10}
\end{equation}
with $H$ being the temperature-salinity gradient ratio $\left( {d\left\langle T \right\rangle {\rm{/}}dz} \right)/\left( {d\left\langle S \right\rangle {\rm{/}}dz} \right)$, $\alpha$ and $\beta$ are the thermal expansion coefficient and the saline contraction coefficient, respectively, which can be calculated from expressions
\begin{equation}
\alpha \left( {T,S} \right) = {\left. {\frac{1}{V}\frac{{\partial V}}{{\partial T}}} \right|_S}
\quad {\rm and} \quad
\beta \left( {T,S} \right) = {\left. {\frac{1}{V}\frac{{\partial V}}{{\partial S}}} \right|_T},
\label{eq13}
\end{equation}
under the assumption of atmospheric pressure. The (specific) volume $V$ has been described by a 75-term polynomial expression in the new version of TEOS-10 standard.
Related formulae and calculations have been reported in \cite{mcdougall2009international}, and developed in the TEOS-10 toolbox \cite{mcdougall2011getting}.
\textit{Based on Eq. (\ref{eq9})-(\ref{eq10}), 
	and using the $\alpha\left(\left<T\right>,\left<S\right>\right)$ and $\beta\left(\left<T\right>,\left<S\right>\right)$ in TEOS-10 toolbox, one can calculate the eddy diffusivity ratio $d_r\left(\left<T\right>,\left<S\right>,H\right)$.}

\begin{figure}
	\centering
	\includegraphics[width=0.32\textwidth]{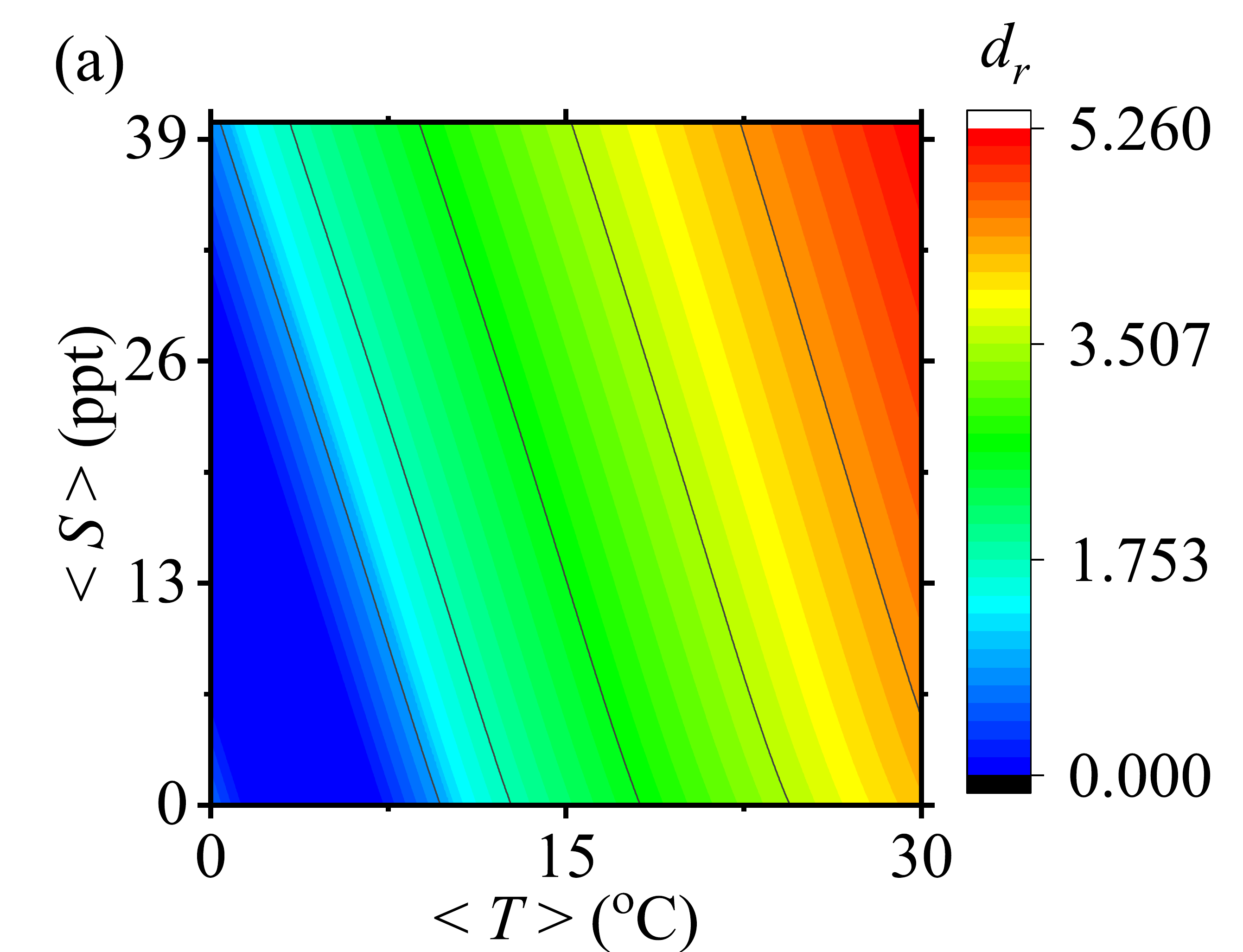}
	\includegraphics[width=0.32\textwidth]{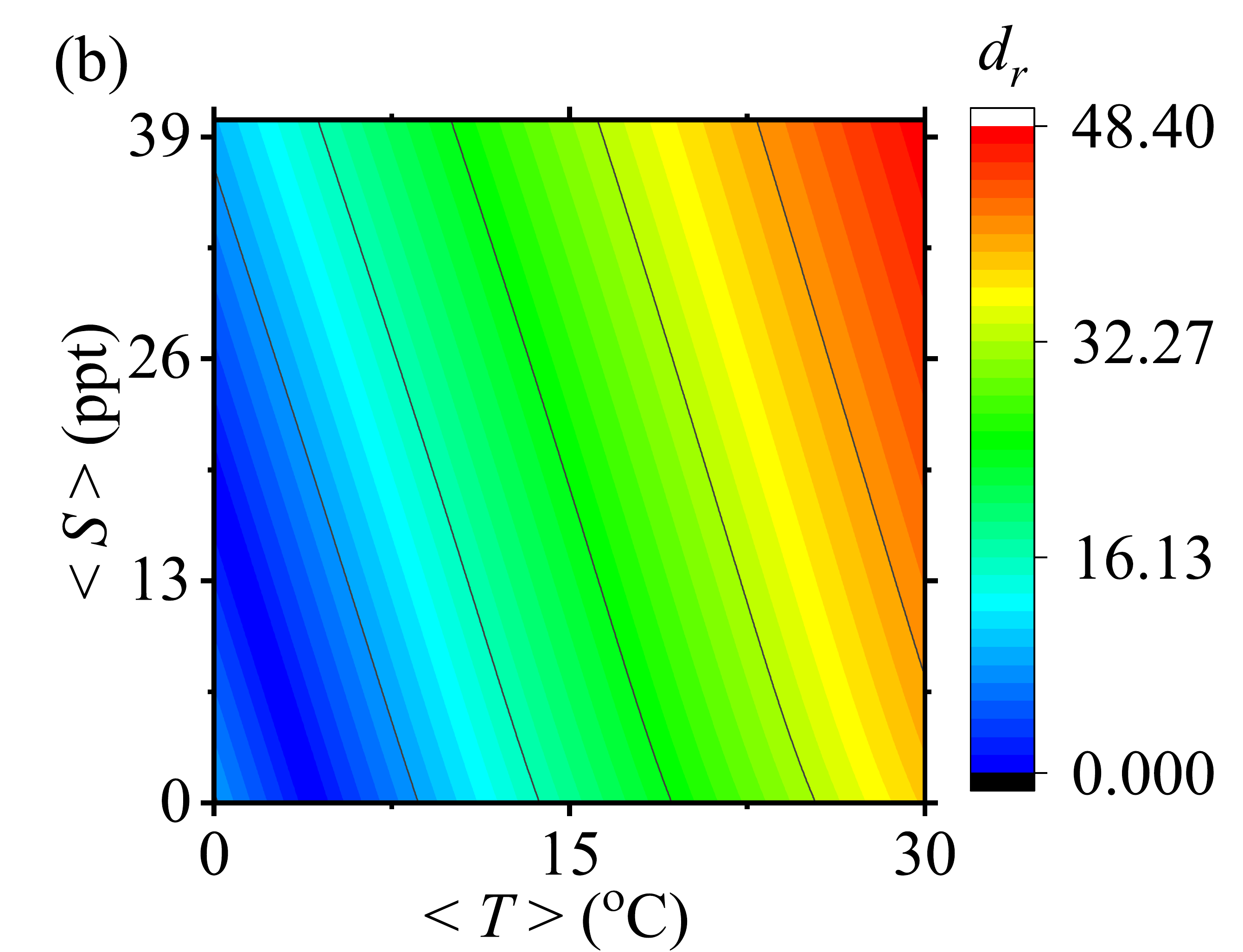}
	\includegraphics[width=0.32\textwidth]{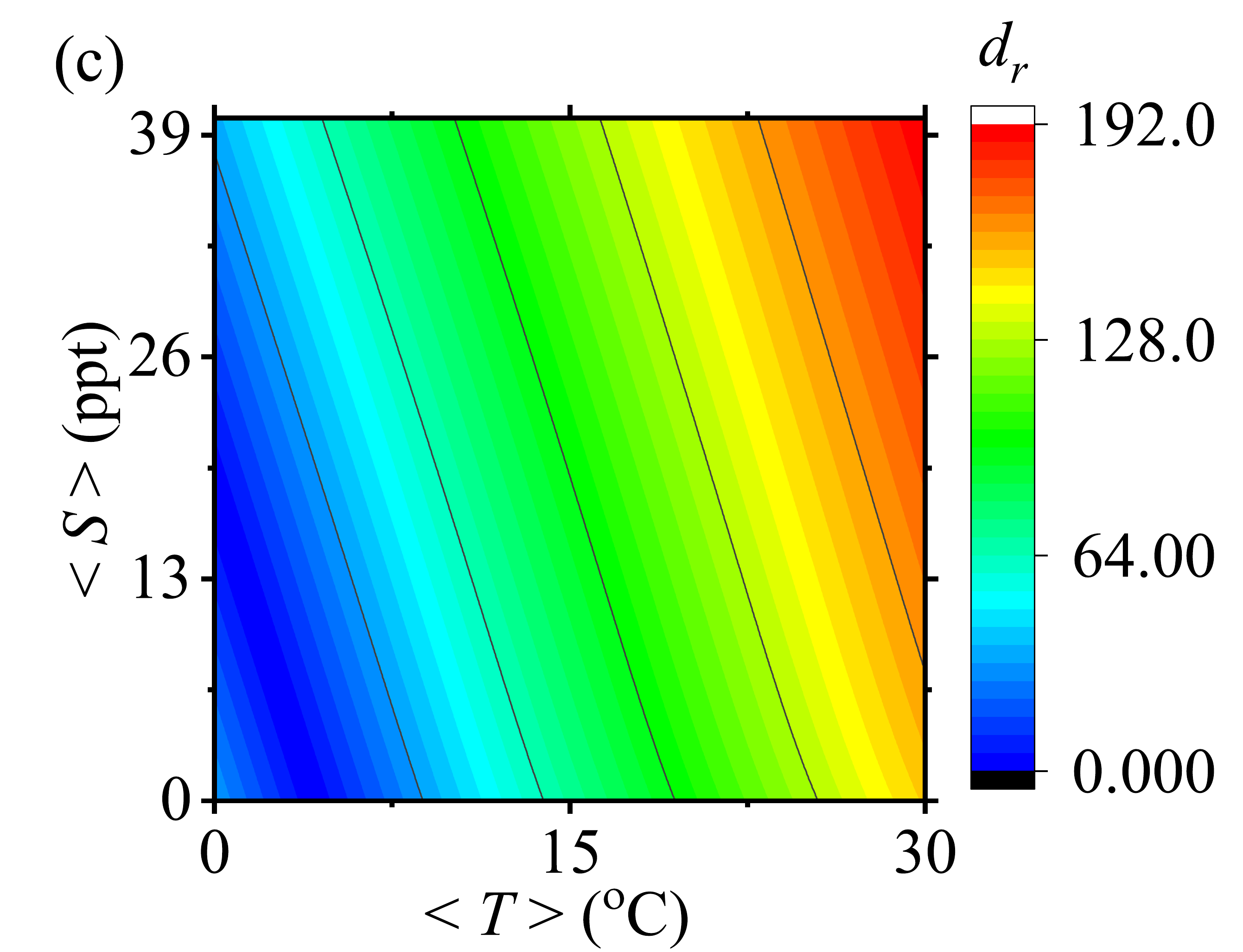}
	\caption{$d_r$ varying with $\left<T\right>$ and $\left<S\right>$ at different values of $H$. (a) $H = -10 ^\circ\rm{C}\cdot \rm{ppt}^{-1}$; (b) $H = -100 ^\circ\rm{C}\cdot \rm{ppt}^{-1}$; (c) $H = -400 ^\circ\rm{C}\cdot \rm{ppt}^{-1}$.}
	\label{fig:dr}
\end{figure}

Based on all the results of this section, Fig. \ref{fig:dr} presents $d_r$ changing with $\left< T \right>$ and $\left< S \right>$, for several fixed values of temperature-salinity gradient ratio $H$. 

\section{Calculation of the linear coefficients of temperature and salinity}
In this section we derive, on the basis of the water refractive-index polynomial of Quan and Fry \cite{QuanFry}, the expressions for linear coefficients to the power spectrum characterizing contributions from temperature and salinity fluctuations which depend on average temperature $\left\langle T\right\rangle [^\circ \rm{C}]$, average salinity $\left\langle S\right\rangle [\rm{ppt}]$ and light wavelength $\lambda [\rm{nm}]$. The polynomial expression obtained in \cite{QuanFry} for the refractive index $n(T, S,\lambda)$ agrees well with Sager's data \cite{sager1974refraktion}  being less than $5\times10^{-5} \rm{1}$. This allows us to assume the validity ranges for the average temperature and average salinity concentration as: $0\ ^\circ \rm{C}\le T\le30\ ^\circ\rm{C}$ and $0\ \rm{ppt}\le S\le40\ \rm{ppt}$, respectively.
Besides, by comparing their model with Austin and Halikas' experimental data \cite{austin1976index}, Quan and Fry have shown its occuracy in interval $400\ \rm{nm} \leq \lambda \leq 700\ \rm{nm}$. The empirically fitted polynomial of \cite{QuanFry} has form:
\begin{equation}
\begin{split}
n( T, S,\lambda)&=a_0+(a_1+a_2 T+a_3 T^2)S +a_4T^2   \\&
+\frac{a_5+a_6 S+a_7 T}{\lambda}+\frac{a_8}{\lambda^2}+\frac{a_9}{\lambda^3},
\label{eq14}
\end{split}
\end{equation}
where constants $a_i$, $i=\{0, ..., 9\}$ have the following values
\begin{equation}
\begin{split}
&a_0=1.31405, \quad a_1=1.779\times 10^{-4}, \quad a_2=-1.05\times 10^{-6},\\&  a_3=1.6\times 10^{-8}, \quad a_4=-2.02 \times 10^{-6}, \quad a_5=15.868, \\& a_6=0.01155, \quad a_7=-0.00423, \quad a_8=-4382, \quad a_9=1.1455\times 10^{6}.
\label{eq15}
\end{split} 
\end{equation}
This model was shown to be consistent with all the data  taken up to then and to be a generalization or correction for previously introduced models (see \cite{QuanFry} and references wherein). 

Let us first represent the refractive index $n$ as a sum of its average value $n_0$ and relative fluctuation $n'$:
\begin{equation}
n=n_0(\langle T \rangle, \langle S \rangle, \lambda) +n', 
\label{eq16}
\end{equation} 
where the latter portion can be approximately linearized as
\begin{equation}
n'\approx A(\langle T \rangle, \langle S \rangle, \lambda)T'+B(\langle T \rangle, \langle S \rangle, \lambda)S',
\label{eq17}
\end{equation}
with $T'$ and $S'$ being the fluctuating components of the temperature and salinity concentration distributions, respectively, while $A$ and $B$ being the linear coefficients. Unlike in the previous oceanic refractive-index spectrum models, essentially all based on approach taken in \cite{Nikishovs},
here $A$ and $B$ are not constants but functions of the water's average temperature, average salinity and and the wavelength of light. In order to determine such functional dependence for $A$ and $B$, we examine the first-order Taylor approximation
\begin{equation}
\begin{split}
dn\left( {T,S,\lambda } \right){\rm{ }} = \frac{{\partial n\left( {T,S,\lambda } \right)}}{{\partial T}}dT + \frac{{\partial n\left( {T,S,\lambda } \right)}}{{\partial S}}dS + \frac{{\partial n\left( {T,S,\lambda } \right)}}{{\partial \lambda }}d\lambda .
\label{eq18}
\end{split}
\end{equation}
Setting $d\lambda=0$ implies that
\begin{equation}
\begin{split}
A\left( {\langle T\rangle ,\langle S\rangle ,\lambda } \right) &= {\left. {\frac{{\partial n\left( {T,S,\lambda } \right)}}{{\partial T}}} \right|_{T = \left\langle T \right\rangle ,S = \left\langle S \right\rangle }}\\&
=a_2\langle S \rangle +2a_3 \langle T \rangle \langle S \rangle +2a_4\langle T \rangle + \frac{a_7}{\lambda},
\label{eq19}
\end{split}
\end{equation}
and 
\begin{equation}
\begin{split}
B\left( {\langle T\rangle ,\langle S\rangle ,\lambda } \right) &= {\left. {\frac{{\partial n\left( {T,S,\lambda } \right)}}{{\partial S}}} \right|_{T = \left\langle T \right\rangle ,S = \left\langle S \right\rangle }}\\&
=a_1+a_2 \langle T \rangle +a_3 \langle T\rangle ^2 +\frac{a_6}{\lambda}.
\label{eq20}
\end{split}
\end{equation}

It can be deduced from Eqs. (\ref{eq19}) and (\ref{eq20}) that at fixed values $\langle T \rangle=20^\circ\rm{C}$, $\langle S \rangle =34.9\rm{ppt}$ and $\lambda = 532\rm{nm}$ the linear coefficients take on values $A=-10.31\times10^{-5}\rm{deg^{-1}1}$ and $B=1.85\times10^{-4}\rm{g^{-1}1}$. These values somewhat differ from the Nikishovs' result developed with the help of formulas in Ref. \cite{Ruddick}: $A=-2.6\times10^{-4}\rm{deg^{-1}1}$ and $B=1.750\times10^{-4}\rm{g^{-1}1}$ but do agree with the widely accepted approximation $A\approx-10^{-4}\rm{deg^{-1}1}$ and $B\approx2\times10^{-4}\rm{g^{-1}1}$ \cite{mobley1994light}.

\textit{Equation (\ref{eq17}) together with Eqs. (\ref{eq19}) and (\ref{eq20}) constitute the main result of this section. They provide approximations to the linear coefficients of temperature and salinity contributions to the natural water's refractive-index fluctuations varying with $\left\langle T\right\rangle $, $\left\langle S\right\rangle $ and $\lambda$.}

\begin{figure}
	\centering
	\includegraphics[width=0.4\textwidth]{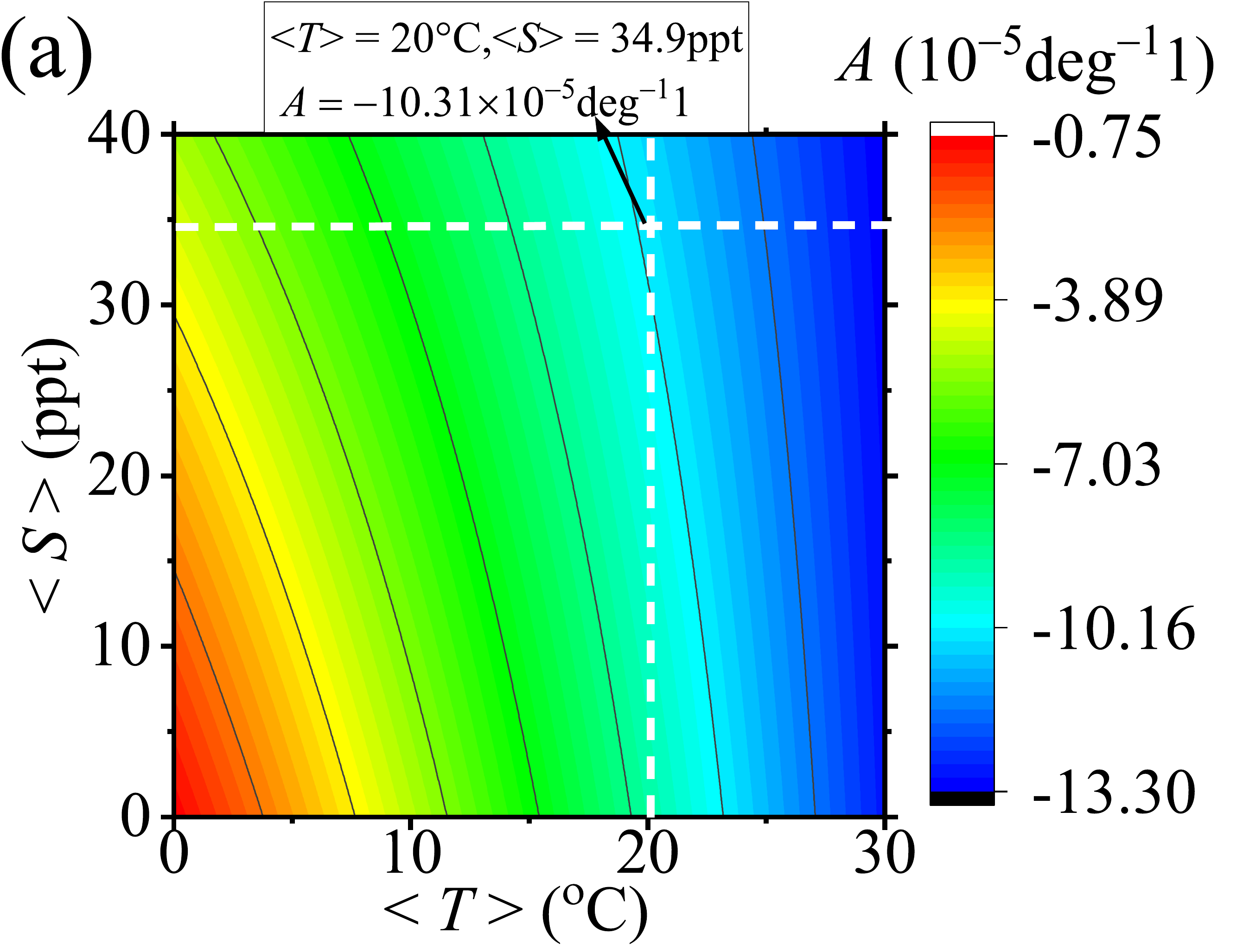}
	\includegraphics[width=0.4\textwidth]{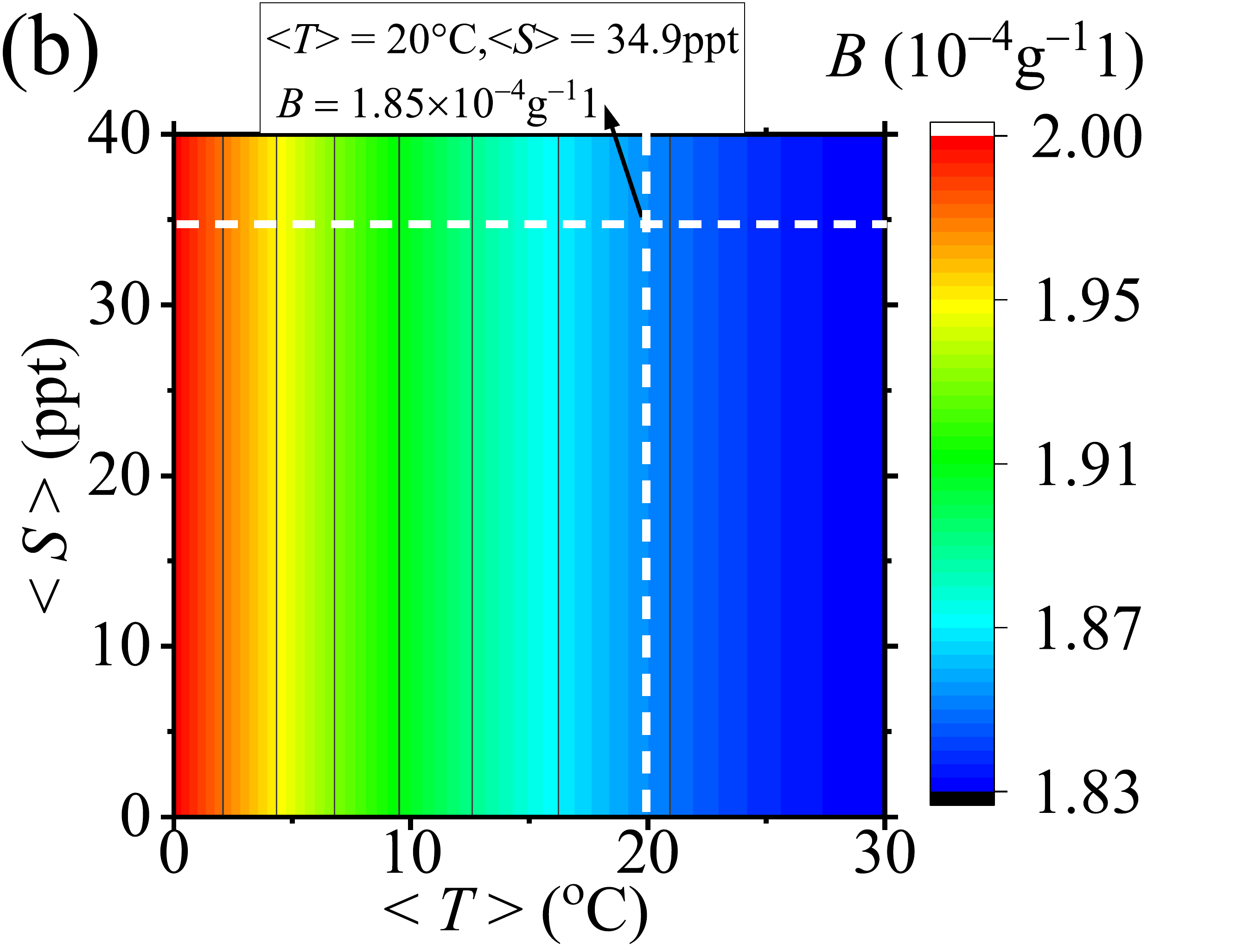}
	\caption{The coefficients $A$ and $B$ varying with $\langle T\rangle$ and $\langle S\rangle$ at $\lambda = 532 nm$.}
	\label{fig:AB_TS}
\end{figure}
\begin{figure}
	\centering
	\includegraphics[width=0.3\textwidth]{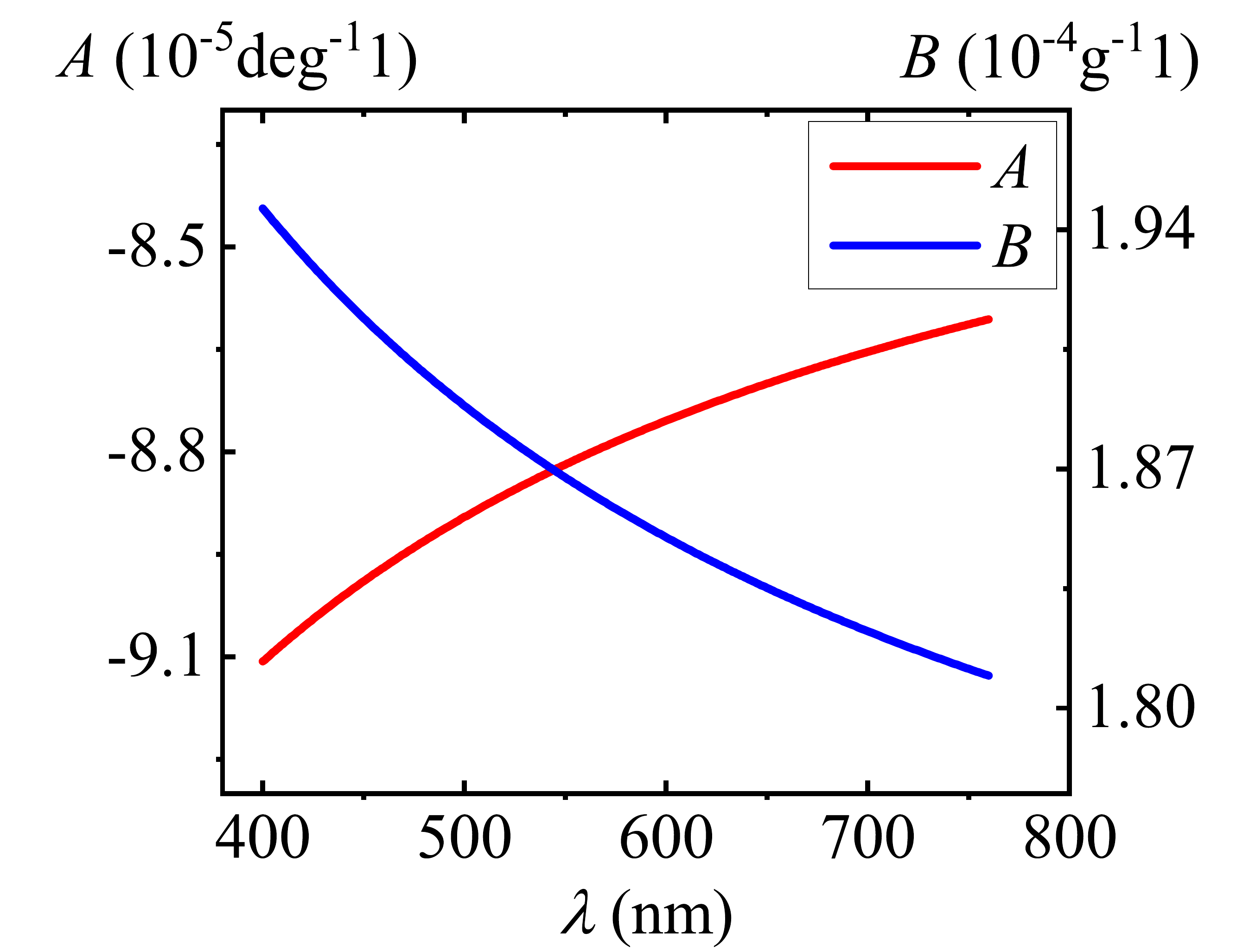}
	\caption{$A$ and $B$ varying with $\lambda$ when $\langle T \rangle = 15 ^\circ C, \langle S \rangle = 34.9 \rm{ppt}$.}
	\label{fig:AB_lambda}
\end{figure}
More details regarding of $A$ and $B$ varying with $\left<T\right>$, $\left<S\right>$ and $\lambda$ are given in Figs. \ref{fig:AB_TS} and \ref{fig:AB_lambda}. Figure \ref{fig:AB_TS} presents variation of $A$ and $B$ as functions of $\left<T\right>$ and $\left<S\right>$ when $\lambda = 532\ \rm{nm}$.
Coefficient $A$ decreases with increasing $\left<T\right>$ and $\left<S\right>$ but $B$ only  decreases with increasing $\left<T\right>$ but does not vary with $\left<S\right>$. Such invariance can also be directly established from Eq.(\ref{eq20}).
Figure \ref{fig:AB_lambda} shows the wavelength dependence of $A$ and $B$ at $\left<T\right>=15\ ^\circ \rm{C}$ and $\left<S\right>=34.9\  \rm{ppt}$.

\begin{figure}
	\centering
	\includegraphics[width=0.32\textwidth]{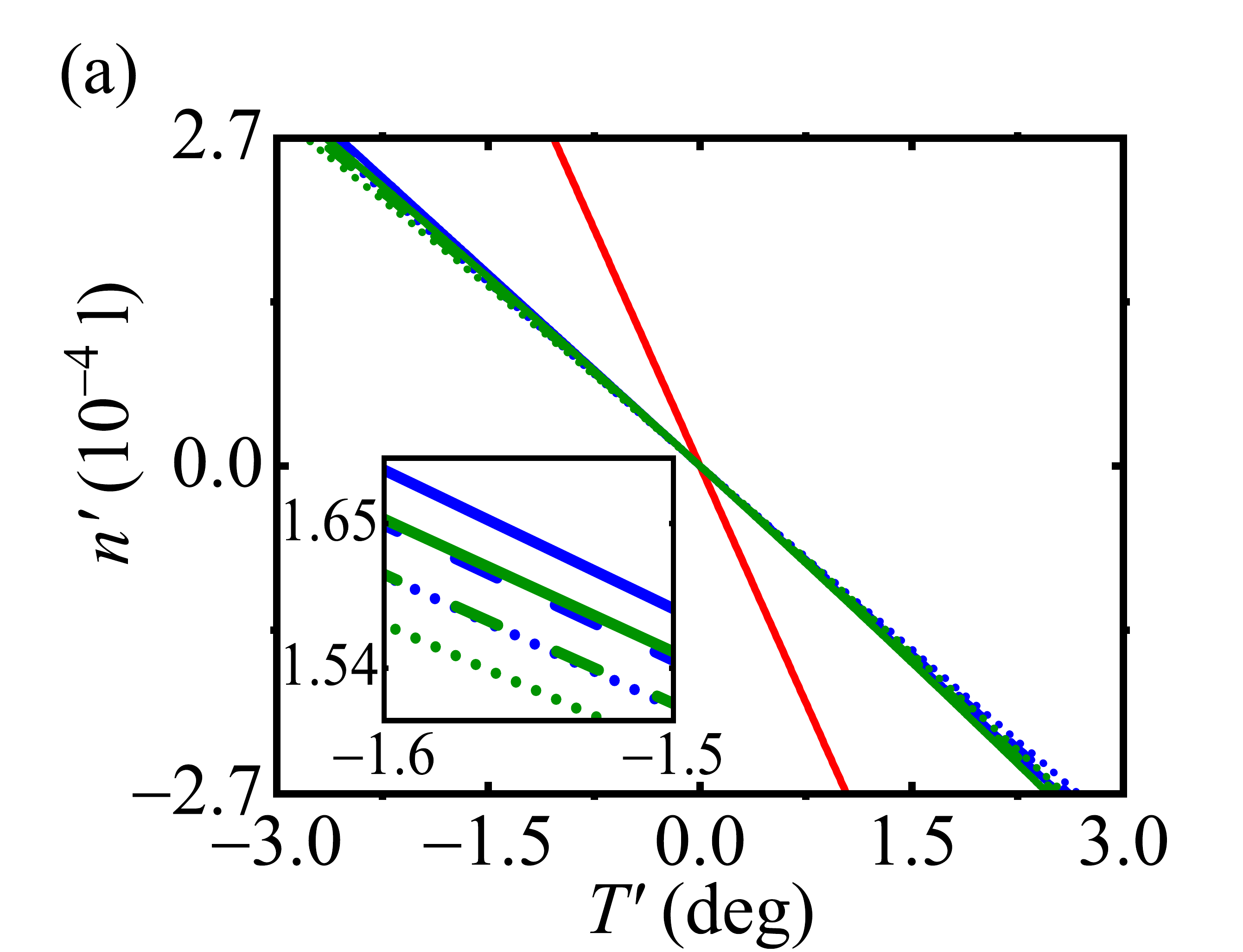}
	\includegraphics[width=0.18\textwidth]{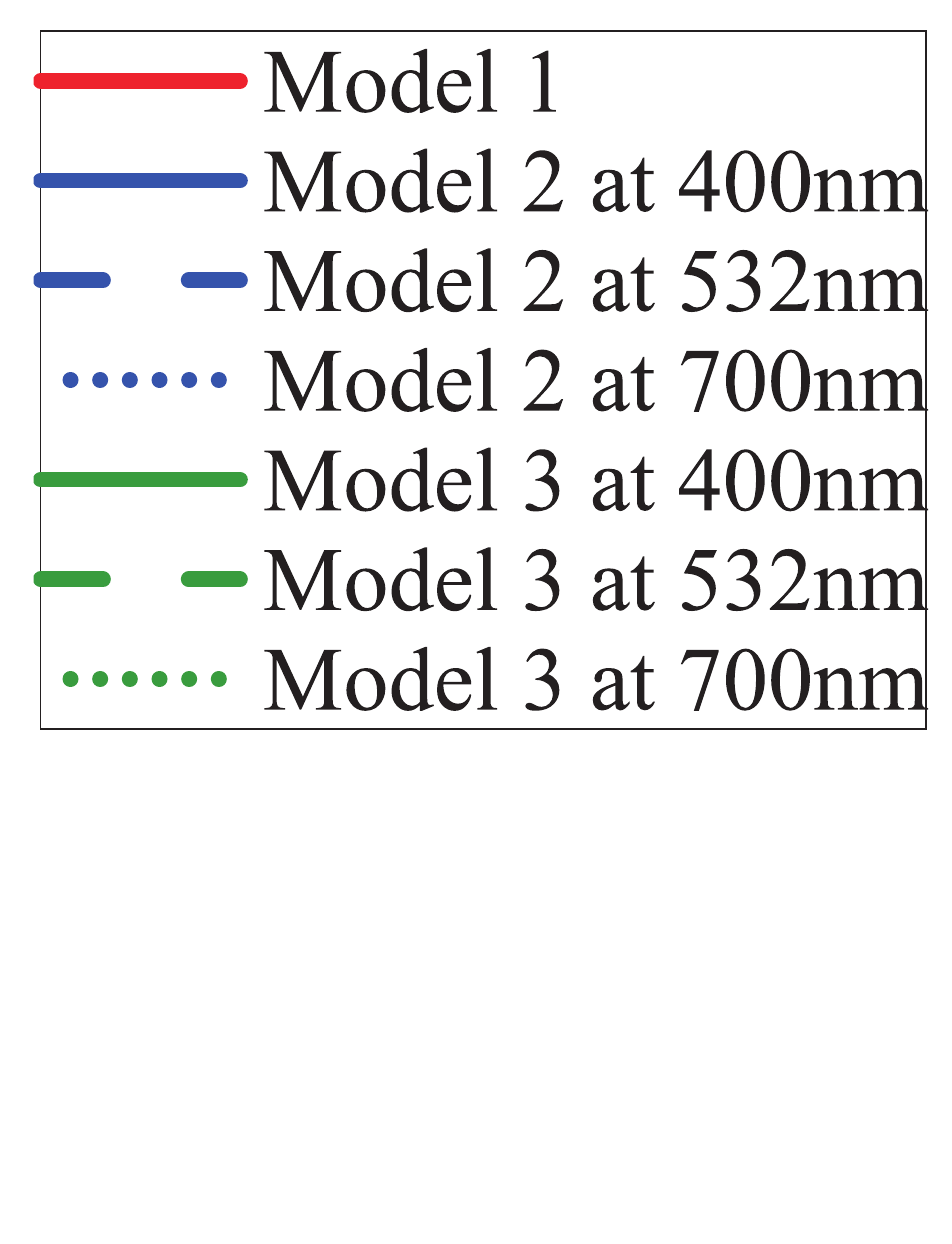}
	\includegraphics[width=0.32\textwidth]{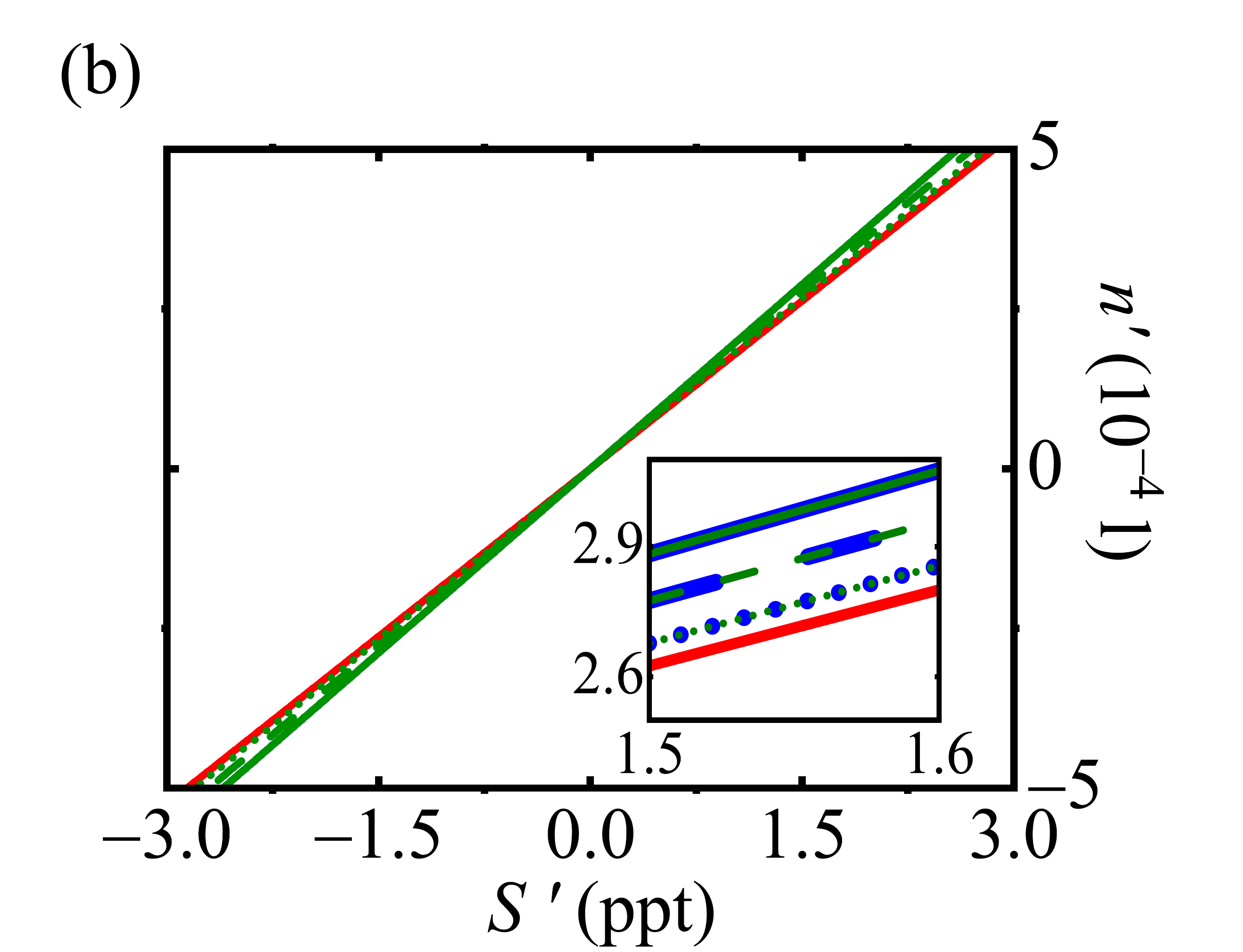}
	\caption{(a) $n'$ varying with $T'$, and (b) $n'$ varying with $S'$ in three models: \\  Model 1: Nikishovs' model; \\ Model 2: Our linear model in Eq. (\ref{eq17}) with Eqs. (\ref{eq18})-(\ref{eq19}); \\ Model 3: Quan and Fry's formula containing full polynomial. }
	\label{fig:n'}
\end{figure}
Figure \ref{fig:n'} compares linear approximation given in Eq. (\ref{eq17}) with the Quan and Fry's formula (see Eq. (\ref{eq14}) of Ref. \cite{QuanFry}) as well as with the Nikishovs' model, by plotting $n'$ varying with $T'$ and $S'$ when $\left<T\right> = 20\ ^\circ\rm{C}$ and $\langle S\rangle = 34.9\ \rm{ppt}$.
It shows that Nikishovs' model obviously deviates from the Quan and Fry's model while our  approximation given by  Eq. (\ref{eq18}) agrees well with it.

\section{Power spectrum for any average temperature and salinity}

In this section we will incorporate the expressions obtained in the previous sections for the Prandtl/Schmidt numbers, the eddy diffusivity ratio and the linear coefficients $A$
 and $B$ into the spatial power spectrum of the water refractive-index fluctuations. According to Ref. \cite{Nikishovs} the power spectrum can be expressed as
\begin{equation}
\begin{split}
{\Phi _{\rm{n}}}(\kappa, \langle T \rangle, \langle S \rangle, \lambda) =& {A^2(\langle T \rangle, \langle S \rangle, \lambda)}{\Phi _{\rm{T}}}(\kappa ) + {B^2(\langle T \rangle, \langle S \rangle, \lambda)}{\Phi _{\rm{S}}}(\kappa ) \\& +  2A(\langle T \rangle, \langle S \rangle, \lambda)B(\langle T \rangle, \langle S \rangle, \lambda){\Phi _{{\rm{TS}}}}(\kappa ),
\label{eq21}
\end{split}
\end{equation}
where ${\Phi _{\rm{T}}}(\kappa )$, ${\Phi _{\rm{S}}}(\kappa )$, and ${\Phi _{{\rm{TS}}}}(\kappa )$ are the temperature spectrum, the salinity spectrum, and the co-spectrum, respectively, and  
$A$ and $B$ are the linear coefficients varying with $<T>$, $<S>$ and $\lambda$ obtained in  Eqs. (\ref{eq19}) and (\ref{eq20})  \cite{footnote2}.

For each of these three spectra, we will apply the analytic fit \cite{YaoKor}   
\begin{equation}
 \begin{split}
{\Phi _i}(\kappa, \langle T \rangle, \langle S \rangle) &= \left[ {1 + {\rm{21}}{\rm{.52}}{{(\kappa \eta )}^{{\rm{0}}{\rm{.63}}}}{c_i}^{{\rm{0}}{\rm{.02}}} - {\rm{18}}{\rm{.17}}{{(\kappa \eta )}^{0.58}}{c_i}^{{\rm{0}}{\rm{.04}}}} \right]\\
	& \times\frac{1}{{4\pi}}\beta {\varepsilon ^{ - \frac{1}{3}}}{\kappa ^{ - \frac{11}{3}}}{\chi _i} \exp \left[ { - 176.41{{(\kappa \eta )}^2}{c_i}^{{\rm{0}}{\rm{.96}}}} \right], \quad i \in \{{\rm{T}},{\rm{S}},{\rm{TS}}\} , 
\label{eq22}
\end{split}
\end{equation}
where the ensemble-averaged variance dissipation rates ${\chi _i} \left( i\in\left\lbrace T,S,TS\right\rbrace \right) $ are defined by \cite{Nikishovs,Elamassie}
\begin{equation}
{\chi _T} = {K_T}{\left( {\frac{{d\left\langle T \right\rangle }}{{dz}}} \right)^2},
\quad {\chi _S} = {K_S}{\left( {\frac{{d\left\langle S \right\rangle }}{{dz}}} \right)^2},
\quad {\chi _{TS}} = \frac{{{K_T} + {K_S}}}{2}\left( {\frac{{d\left\langle T \right\rangle }}{{dz}}} \right)\left( {\frac{{d\left\langle S \right\rangle }}{{dz}}} \right),
\label{eq23}
\end{equation}
the Kolmogorov microscale $\eta [\rm{m}^{-1}]$ is
\begin{equation}
\begin{split}
\eta  &= {\nu ^{3/4}}{\varepsilon ^{ - 1/4}}\\
&= {\left[ {\frac{{\mu (\langle T\rangle ,\langle S\rangle )}}{{\rho (\langle T\rangle ,\langle S\rangle )}}} \right]^{3/4}}{\varepsilon ^{ - 1/4}},
\label{eq24}
\end{split}
\end{equation}
where $\varepsilon$ is the energy dissipation rate $[m^2/s^3]$. In Eq. \eqref{eq22} the non-dimensional parameters $c_i \left( i\in\left\lbrace T,S,TS\right\rbrace \right)$ are
\begin{equation}
\begin{split}
&c_T= 0.072^{4/3}\beta {Pr}^{-1}(\langle T \rangle, \langle S \rangle),  \quad c_S= 0.072^{4/3}\beta {Sc}^{-1} (\langle T \rangle, \langle S \rangle), \\&
c_{TS}=0.072^{4/3}\beta \frac{Pr(\langle T \rangle, \langle S \rangle)+Sc(\langle T \rangle, \langle S \rangle)}{2Pr(\langle T \rangle, \langle S \rangle)Sc(\langle T \rangle, \langle S \rangle)},
\label{eq25}
\end{split}
\end{equation}
$c_{TS}$ is based on the coupling between $Pr$ and $Sc$ \cite{Yao,Yi}; $K_T$ and $K_S$, as before, are the eddy diffusivity of temperature and salinity, respectively.
Combining Eq. (\ref{eq23}) with Eqs. (\ref{eq9})-(\ref{eq13}), we get
\begin{equation}
\begin{split}
&{\chi _S} \left( \left\langle T \right\rangle, \left\langle S \right\rangle, H, \chi _T\right) =\frac{{{d_r\left( \left\langle T \right\rangle, \left\langle S \right\rangle, H\right) }}}{{{H^2}}}{\chi _T}, \\&{\chi _{TS}}\left( \left\langle T \right\rangle, \left\langle S \right\rangle, H, \chi _T\right) =\frac{{1 + {d_r\left( \left\langle T \right\rangle, \left\langle S \right\rangle, H\right) }}}{{2H}}{\chi _T}.
\label{eq26}
\end{split}
\end{equation}
where, as shown in Section 3,  $H$ is the temperature-salinity gradient ratio, and $d_r$ can be directly calculated from $\left\langle T \right\rangle$, $\left\langle S \right\rangle$ and $H$.

\textit{The power spectrum model given by Eqs. (\ref{eq21}) - (\ref{eq26}) is the main result of our study. In combination with the results of Sections 2, 3 and 4, it gives the 2nd-order analytic description of the natural water optical turbulence with the wide-range average temperatures and salinity concentrations occurring in the Earth's oceans, seas, bays, rivers and lakes, at any geographic region, under a variety of meteorological conditions, and for all visible wavelengths}.  

\begin{figure}
	\centering
	\includegraphics[width=0.3\textwidth]{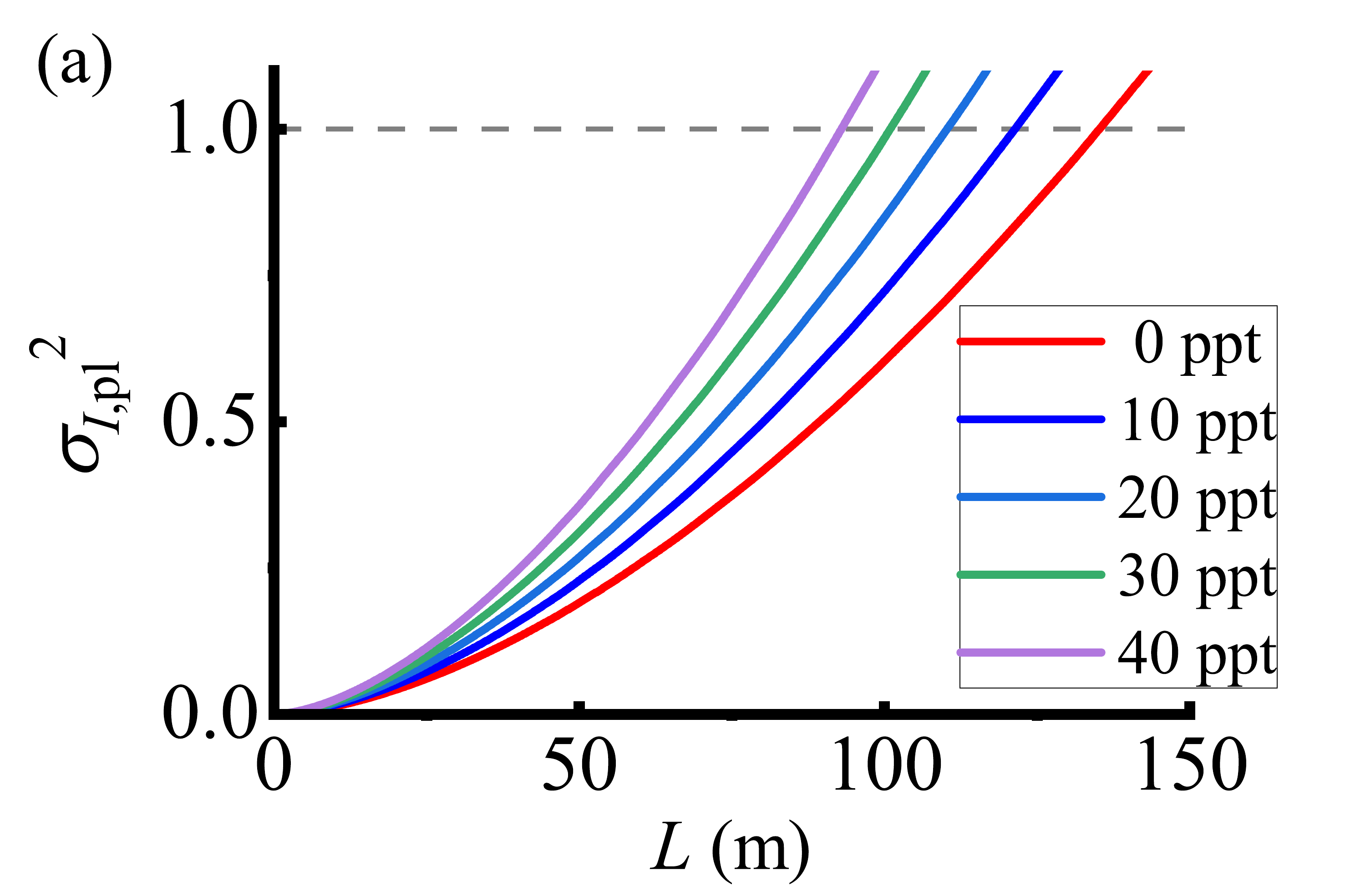}
	\includegraphics[width=0.3\textwidth]{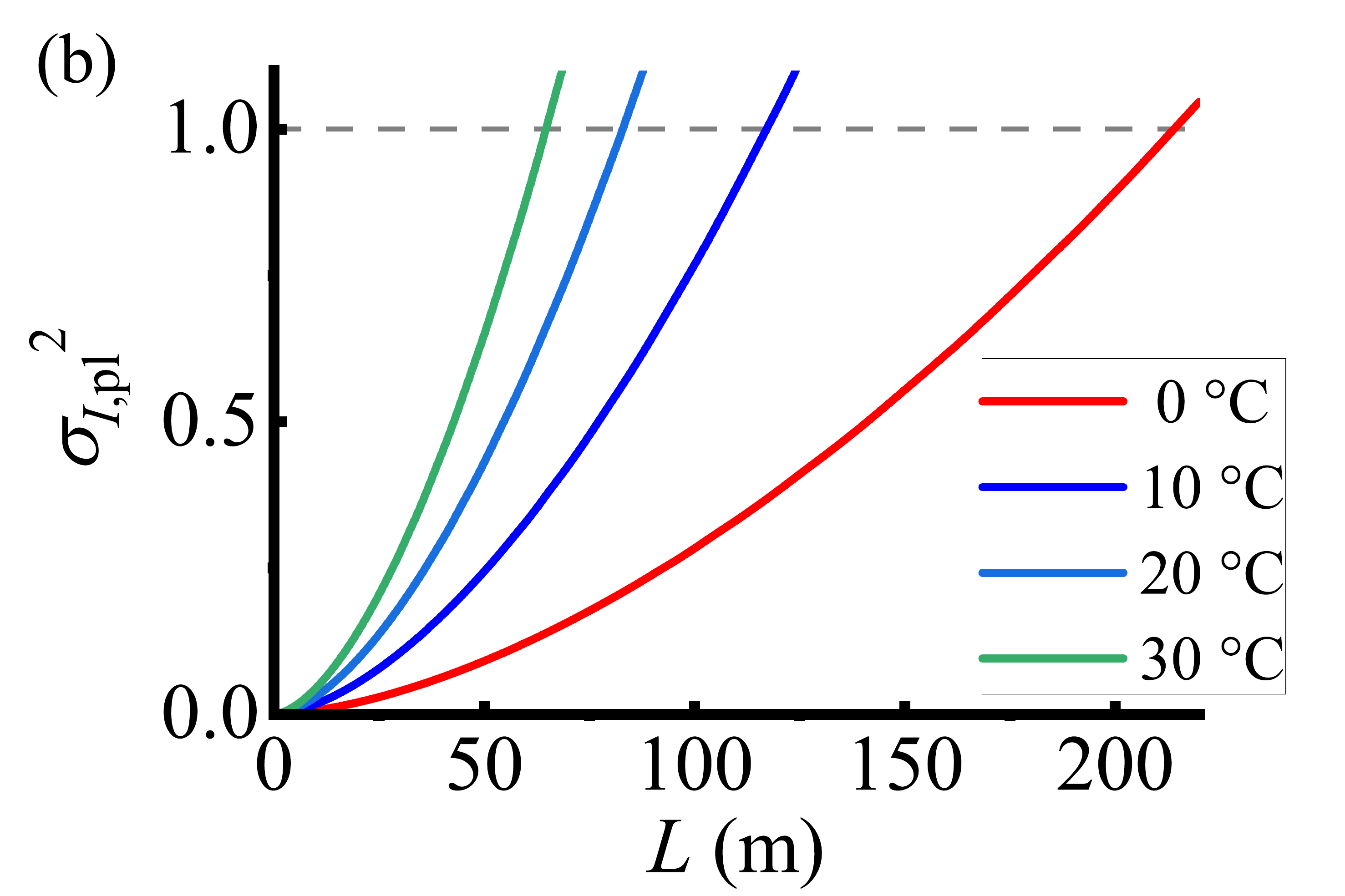}
	\includegraphics[width=0.3\textwidth]{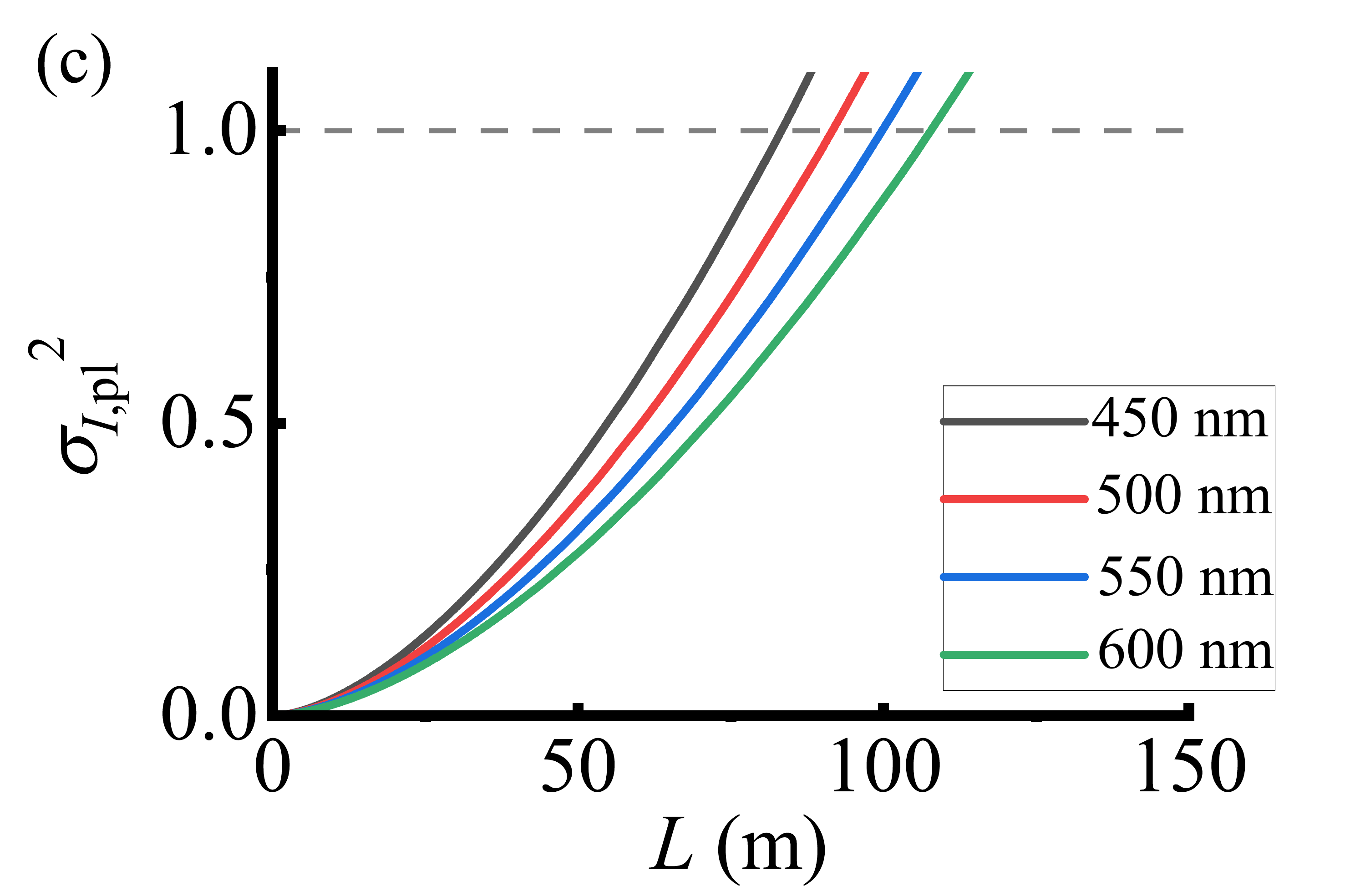}
	\caption{The scintillation index $\sigma _{I,{\rm{pl}}}^2(L)$ with different values of (a) average salinity, (b) average temperature, and (c) wavelength.}
	\label{fig:Sci}
\end{figure}

\section{Light scintillation in natural waters}

In this section we will explore the effects of the average temperature, the average salinity concentration and the wavelength entering the developed power spectrum model $\Phi_n(\kappa,\langle T \rangle, \langle S \rangle, \lambda)$ on the scintillation index of a plane wave, also known as the Rytov variance \cite{AP}.  It is given by expression
\begin{equation}
\begin{split}
\sigma _{I,{\rm{pl}}}^2(L) &= \frac{{8{\pi ^2}{k_0^2}L}}{{n_0^2}}\int_0^1 {d\xi \int_0^\infty  { \kappa \Phi_n(\kappa,\langle T \rangle, \langle S \rangle, \lambda)\left[ {1 - \cos \left( {\frac{{L{\kappa ^2}\xi }}{k_0}} \right)} \right] d\kappa} }\\
&= \frac{{8{\pi ^2}k_0^2L}}{{n_0^2}}\int_0^\infty  {\kappa \Phi_n(\kappa,\langle T \rangle, \langle S \rangle, \lambda)\left[ {1 - \frac{{\sin \left( {L{\kappa ^2}{\rm{/}}{k_0}} \right)}}{{L{\kappa ^2}{\rm{/}}{k_0}}}} \right]d\kappa}, \label{eq27}
\end{split}
\end{equation} 
where $n_0$ is the average value equal to $n\left( {\left\langle T \right\rangle ,\left\langle S \right\rangle , \lambda  } \right)$,  $k_0=2\pi/\lambda$ is the wavenumber, and $L$ is the propagation distance. This quantity is one of the most crucial observables of optical turbulence and is frequently used for separation of weak and strong turbulence regimes \cite{AP}. It was used in Ref. \cite{SI} for the very first analysis of optical scintillation underwater.

Figure \ref{fig:Sci} shows scintillation index $\sigma _{I,{\rm{pl}}}^2(L)$ for several fixed values of $\left<S\right>$, $\left<T\right>$ and $ \lambda $ and the following fixed values of parameters: $\varepsilon = 10^{-2}\rm{m^2s^{-3}}$, $H =-20\ ^\circ\rm{C}\cdot \rm{ppt}^{-1}$ and $\chi_T=10^{-5}\rm{K^2s^{-1}}$. We also set $\left<T\right> = 15\ ^\circ\rm{C}$ and $\left<S\right> = 34.9\ \rm{ppt}$ in Fig. \ref{fig:Sci}(a); $\left<T\right> = 15\ ^\circ\rm{C}$ and $\lambda=532\ \rm{nm}$ in Fig. \ref{fig:Sci}(b); $\left<S\right> = 34.9\ \rm{ppt}$ and $\lambda=532\ \rm{nm}$ in Fig. \ref{fig:Sci}(c).
It is shown that \textit{larger $\left<S\right>$ and/or $\left<T\right>$ lead to stronger effects of turbulence on the plane wave and result in a larger scintillation \cite{footnote4,footnote3}. We also conclude that the shorter the wavelength of light the stronger the scintillations are.} 

\begin{figure}
	\centering
	\includegraphics[width=0.35\textwidth]{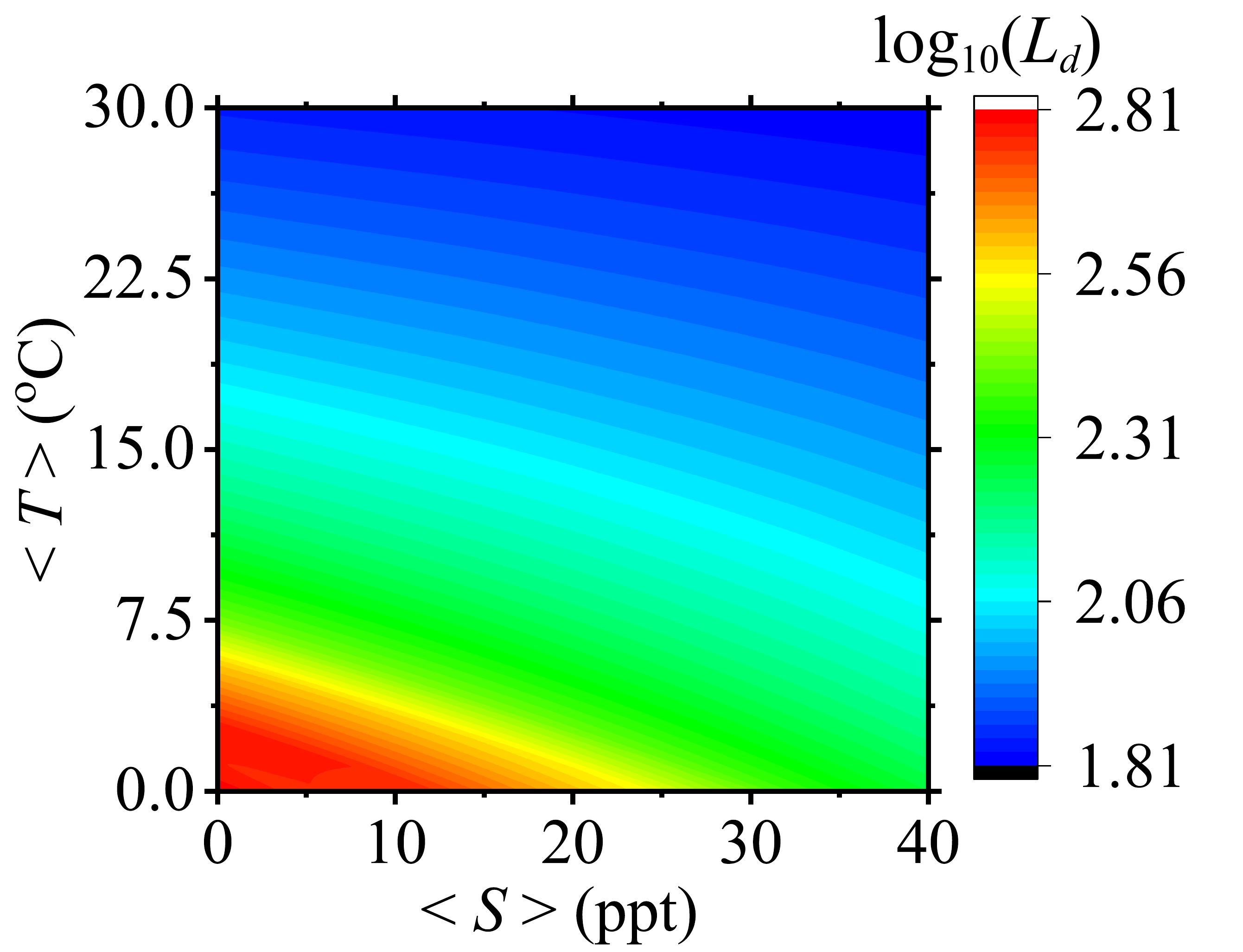}
	\caption{$\log_{10}{\left( L_d\right) }$ varying with $\left<T\right>$ and $\left<S\right>$ where $L_d$ has the unit [m].}
	\label{fig:Ldistance}
\end{figure}

In addition, on solving equation
\begin{equation}
\sigma _{I,{\rm{pl}}}^2({L_d}) = 1,
\label{eq28}
\end{equation}
we can find the threshold distance $L_d$ between weak  ($L \ll {L_d}$) and strong ($L \gg {L_d}$) turbulence regimes. The density plot of $\log_{10}{\left[ L_d\left(\left<T\right>,\left<S\right>\right)\right] }$ is illustrated in Fig. \ref{fig:Ldistance} with $\lambda=532\rm{nm}$, $\varepsilon = 10^{-2}\rm{m^2s^{-3}}$, $H =-20\ ^\circ\rm{C}\cdot \rm{ppt}^{-1}$ and $\chi_T=10^{-5}\rm{K^2s^{-1}}$.
\textit{It is evident that larger values of $L_d$ correspond to lower $\left<T\right>$ and $\left<S\right>$} which is in agreement with Fig. \ref{fig:Sci}(a)-(b).

\section{Summary and conclusion}
We have derived the expressions for the Prandtl/Schmidt numbers, the eddy diffusivity ratio varying with average temperature $\langle T \rangle$ and average salinity $\langle S \rangle$,
as well as the coefficients $A$ and $B$ of the linear approximation of temperature and salinity contributions to the natural water power spectrum as functions of $\langle T \rangle$, $\langle S \rangle$ and wavelength $\lambda$.
We have found the following:
\begin{itemize}
	\item Larger values of $\langle T \rangle$ or/and smaller values of $\langle S \rangle$ lead to  smaller values of the Prandtl/Schmidt number;
	\item Eddy diffusivity ratio $d_r$ increases with increasing $\langle T \rangle$ and/or $\langle S \rangle$;
	\item Linear coefficient $A$ decreases with increasing $\langle T \rangle$ and/or $\langle S \rangle$ while linear coefficient $B$ decreases with increasing $\langle T\rangle$ but does not vary with $\langle S\rangle$;
	\item A larger wavelength $\lambda$ leads to a larger $A$ and a smaller $B$.
\end{itemize}

Using these results, we have obtained a model of the oceanic optical turbulence spectrum varying with environmental parameters, and have used this model to calculate the scintillation index of a plane wave. Based on the generic numerical calculations we concluded that a larger $\langle T \rangle$, a larger $\langle S \rangle$ and a smaller $\lambda$ would lead to stronger scintillations.

The proposed power spectrum can be used in numerical calculations relating to light propagation in natural turbulent waters with practically any average temperature and average salinity concentration  present in the Earth's boundary layer and with any visible wavelength.

\section*{Appendix: Related thermodynamic parameters varying with $\left<T\right>$ and $\left<S\right>$}
\subsection*{I. The specific heat varying with $\langle T\rangle$ and $\langle S\rangle$}
According to \cite{Jamieson} and \cite{Nayar}, at atmospheric pressure, the specific heat $c_p \left[\rm{J\cdot kg^{-1}\cdot K^{-1}}\right]$ is
\begin{equation}
c_p=1000\times (a_{11}+a_{12}\langle T \rangle+a_{13}\langle T \rangle^2+a_{14}\langle T \rangle^2),
\label{eq29}
\end{equation} 
where 
\begin{equation}
\begin{split}
&a_{11}=5.328-9.76\times10^{-2}\langle S \rangle + 4.04\times 10^{-4}\langle S \rangle^2, \\&
a_{12}=-6.913\times10^{-3}+7.351\times 10^{-4}\langle S \rangle-3.15\times 10^{-6}\langle S \rangle^2, \\&
a_{13}=9.6\times 10^{-6}-1.927 \times 10^{-6}\langle S \rangle +8.23 \times10^{-9}\langle S \rangle^2, \\&
a_{14}=2.5\times 10^{-9}+1.666 \times 10^{-9}\langle S \rangle -7.125 \times10^{-12}\langle S \rangle^2.
\label{eq30}
\end{split}
\end{equation}

\subsection*{II. The thermal conductivity varying with $\langle T\rangle$ and $\langle S\rangle$}
Thermal conductivity $\sigma_T \left[\rm{W\cdot m^{-1}K^{-1}}\right]$ can be calculated using results of  Ref. \cite{ThC}:
\begin{equation}
\begin{split}
\log\left(\sigma_T\right)=&0.434\times\left(2.3-\frac{343.5+0.037 \langle S_h \rangle}{\langle T_h \rangle +273.15}\right) \left[1-\frac{\langle T_h \rangle+273.15}{647.3+0.03\langle S_h \rangle}\right]^{1/3} \\& + {\log\left(240+0.0002 \langle S_h \rangle\right)} - 3,
\label{eq31}
\end{split}
\end{equation}
where 
\begin{equation}
\langle T_h \rangle=1.00024\langle T \rangle, \quad \langle S_h \rangle =\langle S \rangle /1.00472.
\label{eq32}
\end{equation}

\subsection*{III. The dynamic viscosity varying with $\langle T\rangle$ and $\langle S\rangle$}
At atmospheric pressure the dynamic viscosity $\mu \left[\rm{N\cdot s\cdot m^{2}}\right]$ can be finely evaluated using a model of Ref.  \cite{DV1}, obtained as the analytic fit of data reported in \cite{DV1}-\cite{DV4}, where:
\begin{equation}
\mu=\mu_0(a_{21}\langle s \rangle + a_{22} \langle s \rangle^2), \quad \langle s \rangle=\langle S \rangle \times10^{-3}.
\label{eq33}
\end{equation}
where
\begin{equation}
\begin{split}
a_{21}=&1.5409136040+1.9981117208\times 10^{-2}\langle T \rangle -9.5203865864\times 10^{-5}\langle T \rangle^2,  \\
a_{22}=&7.9739318223-7.5614568881\times 10^{-2}\langle T \rangle +4.7237011074 \times 10^{-4}\langle T \rangle^2,  \\&
\label{eq34}
\end{split}
\end{equation}
and
\begin{equation}
\begin{split}
\mu_0=&\left[0.15700386464 \times \left(\langle T \rangle + 64.992620050\right)^2 \right. \\& \left. -91.296496657 \right]^{-1} + 4.2844324477 \times 10^{-5}.
\label{eq35}
\end{split}
\end{equation}

\subsection*{IV. The density of water varying with $\langle T\rangle$ and $\langle S\rangle$}
The density of water $\rho \left[\rm{kg\cdot m^3}\right]$  at atmospheric pressure was fitted in Ref. \cite{DV1} based on data of Refs. \cite{density2} and \cite{density3}.
\begin{equation}
\rho=\rho_T+\rho_S
\label{eq36}
\end{equation}
where temperature-only contribution $\rho_T$ and its adjustment by salinity $\rho_{TS}$ are approximated by polynomials:
\begin{equation}
\begin{split}
\rho_T=& 9.9992293295\times 10^2+2.0341179217\times 10^{-2} \langle T \rangle -6.1624591598 \times 10^{-3}\langle T \rangle^2 \\&+2.2614664708 \times 10^{-5}\langle T \rangle^3 -4.6570659168\times 10^{-8} \langle T \rangle^4,
\label{eq37}
\end{split}
\end{equation}
\begin{equation}
\begin{split}
\rho_{TS}=&\langle s \rangle [8.0200240891 \times 10^2-2.0005183488 \langle T \rangle  +1.6771024982 \times 10^{-2}\langle T \rangle^2\\&-3.0600536746 \times 10^{-5}\langle T \rangle^3 -1.6132224742 \times 10^{-5} \langle T \rangle^2 \langle s \rangle],
\label{eq38}
\end{split}
\end{equation}
where, as before, $\langle s \rangle=10^{-3}\times \langle S \rangle$.

\section*{Acknowledgement}
We thank John Lienhard and his co-workers for developing the website http://web.mit.edu/seawater where the toolbox 'SEAWATER THERMOPHYSICAL PROPERTIES LIBRARY' is provided. We also thank the SCOR/IAPSO Working Group for developing the website http://www.teos-10.org/ and the TEOS-10 toolbox.

\section*{Disclosures}
The authors declare no conflicts of interest.


\begin{thebibliography}{10}
	\newcommand{\enquote}[1]{``#1''}
	
	\bibitem{KorOTreview}
	O.~Korotkova, \enquote{Light Propagation in a Turbulent Ocean,} in
	Progress in Optics, Ed. T. D. Visser, \textbf{64}, 1-43 (Elsevier, 2018).

       \bibitem{Thorpe}
       S.~A.~Thorpe, \textit{The Turbulent Ocean,} 
       (Cambridge: Cambridge University Press, 2007).

        \bibitem{KorRB}
	O.~Korotkova, \textit{Random Beams: Theory and Applications,} 
	(CRC Press, 2013).

        \bibitem{Hill1}
	R.~J. Hill, \enquote{Models of the scalar spectrum for turbulent advection,}
	{\protect\JournalTitle{Journal of Fluid Mechanics}} \textbf{88}, 541–562
	(1978).
	
	\bibitem{Hill2}
	R.~J. Hill, \enquote{Optical propagation in turbulent water,}
	{\protect\JournalTitle{J. Opt. Soc. Am.}} \textbf{68}, 1067--1072 (1978).
	
	\bibitem{Nikishovs}
	V.~V. Nikishov and V.~I. Nikishov, \enquote{Spectrum of turbulent fluctuations
		of the sea-water refraction index,} {\protect\JournalTitle{International
			Journal of Fluid Mechanics Research}} \textbf{27}, 82--98 (2000).
	
	\bibitem{Ruddick}
	B.~Ruddick and T.~Shirtcliffe, \enquote{Data for double diffusers: Physical
		properties of aqueous salt-sugar solutions,} {\protect\JournalTitle{Deep Sea
			Research Part A. Oceanographic Research Papers}} \textbf{26}, 775 -- 787
	          (1979).
	
	
	\bibitem{Elamassie}
	M.~Elamassie, M.~Uysal, Y.~Baykal, M.~Abdallah, and K.~Qaraqe, \enquote{Effect
		of eddy diffusivity ratio on underwater optical scintillation index,}
	{\protect\JournalTitle{J. Opt. Soc. Am. A}} \textbf{34}, 1969--1973 (2017).
	
	\bibitem{Yao}
	J.~Yao, Y.~Zhang, R.~Wang, Y.~Wang, and X.~Wang, \enquote{Practical
		approximation of the oceanic refractive index spectrum,}
	{\protect\JournalTitle{Opt. Express}} \textbf{25}, 23283--23292 (2017).
	
	\bibitem{Yi}
	X.~Yi and I.~B. Djordjevic, \enquote{Power spectrum of refractive-index
		fluctuations in turbulent ocean and its effect on optical scintillation,}
	{\protect\JournalTitle{Opt. Express}} \textbf{26}, 10188--10202 (2018).
	
	\bibitem{YaoKor}
	J.~Yao, H.~Zhang, R.~Wang, J.~Cai, Y.~Zhang, and O.~Korotkova,
	\enquote{Wide-range prandtl/schmidt number power spectrum of optical
		turbulence and its application to oceanic light propagation,}
	{\protect\JournalTitle{Opt. Express}} \textbf{27}, 27807--27819 (2019).
	
	\bibitem{YaoKorLIDAR}
	O.~Korotkova and J.-R. Yao, \enquote{Bi-static lidar systems operating in the
		presence of oceanic turbulence,} {\protect\JournalTitle{Optics
			Communications}} \textbf{460}, 125119 (2020).
	
	\bibitem{QuanFry}
	X.~Quan and E.~S. Fry, \enquote{Empirical equation for the index of refraction
		of seawater,} {\protect\JournalTitle{Appl. Opt.}} \textbf{34}, 3477--3480
	(1995).

	\bibitem{PopeFry} 
	R.~ M.~ Pope and E.~S. Fry, \enquote{Absorption spectrum (380–700 nm) of pure water. II.     Integrating cavity measurements,} {\protect\JournalTitle{Appl. Opt.}} \textbf{36}, 8710--8723 (1997).
	
	\bibitem{poisson1983}
	A.~Poisson and A.~Papaud, \enquote{Diffusion coefficients of major ions in
		seawater,} {\protect\JournalTitle{Marine Chemistry}} \textbf{13}, 265 -- 280
	(1983).

	\bibitem{footnote1}
	As described in \cite{Elamassie}, eddy diffusivity ratio $d_r$ is not
		equal to 1, and it is a piecewise function of density ratio $R_\rho$ which
		varies with thermal expansion coefficient $\alpha$ and saline contraction
		coefficient $\beta$. Here we use TEOS-10 toolbox to calculate $\alpha$ and
		$\beta$ varying with $<T>$ and $<S>$. In combining with $d_r(R_\rho)$, we
		will get details about $d_r$ varying with $<T>$ and $<S>$.
	
	\bibitem{mcdougall2009international}
	T.~McDougall, R.~Feistel, F.~Millero, D.~Jackett, D.~Wright, B.~King,
	G.~Marion, C.~Chen, P.~Spitzer, and S.~Seitz, \enquote{The international
		thermodynamic equation of seawater 2010 (teos-10): Calculation and use of
		thermodynamic properties,} {\protect\JournalTitle{Global Ship-based Repeat
			Hydrography Manual, IOCCP Report No}} \textbf{14} (2009).
	
	\bibitem{mcdougall2011getting}
	T.~J. McDougall and P.~M. Barker, \enquote{Getting started with teos-10 and the
		gibbs seawater (gsw) oceanographic toolbox,}
	{\protect\JournalTitle{SCOR/IAPSO WG}} \textbf{127}, 1--28 (2011).
	
	\bibitem{sager1974refraktion}
	G.~Sager, \enquote{Zur refraktion von licht im meerwasser,}
	{\protect\JournalTitle{Beitr. Meeresk.}} \textbf{33}, 63--72 (1974).
	
	\bibitem{austin1976index}
	R.~W. Austin and G.~Halikas, \textit{The index of refraction of seawater} (UC San
	Diego: Library – Scripps Digital Collection, 1976).
	
	\bibitem{mobley1994light}
	C.~D. Mobley, \textit{Light and Water: Radiative Transfer in Natural Waters}
	(Academic press, 1994).
	
	\bibitem{footnote2}
	The change of refractive index with temperature and salinity is
		nonlinear, which means the linear coefficients of temperature and salinity
		should vary with $<T>$ and $<S>$, and as we know, refractive index changes
		with wavelength. For above reasons, we consider the linear coefficients $A$ 
		and $B$ environment-dependent. We have derived this in section 4. This result 
		is a key point that different from traditional models like \cite{Yao} and
		\cite{YaoKor}.

	\bibitem{AP} L.~C.~Andrews and R.~L.~Phillips, 
        \textit{Laser Beam Propagation in Random Media} (SPIE Press, 2005).

	\bibitem{SI} O.~Korotkova, N.~Farwell and E.~Shchepakina, 
        \enquote{Light scintillation in oceanic turbulence}, 
        {\protect\JournalTitle{Waves in Random and Complex Media}} \textbf{22}, 260--266 (2012). 
	
	\bibitem{footnote4}
	Note that due to natural water absorption most of the power will be absorbed at distances much shorter than those given in Fig. \ref{fig:Sci}. However, in the latest communication technologies some of the commercially available underwater transmission links do operate at the ranges up to 200m (e.g. \cite{ADD1,ADD2,ADD3}).
	
	\bibitem{ADD1}
	C.~{Pontbriand}, N.~{Farr}, J.~{Ware}, J.~{Preisig}, and H.~{Popenoe},
	\enquote{Diffuse high-bandwidth optical communications,} in \emph{OCEANS
		2008,}  (2008), pp. 1--4.
	
	\bibitem{ADD2}
	W. Liu, Z. Xu, and L. Yang, \enquote{SIMO detection schemes for underwater optical wireless communication under turbulence,}
	{\protect\JournalTitle{Photon. Res.}} \textbf{3}, 48--53 (2015).
	
	\bibitem{ADD3}
	N. Saeed, A. Celik, T. Y. Al-Naffouri, and M.-S. Alouini, \enquote{Underwater optical wireless communications, networking, and localization: A survey,}
	{\protect\JournalTitle{Ad Hoc Networks}} \textbf{94}, 101935 (2019).
	
	\bibitem{footnote3}
          The positive correlation between the plane wave scintillation and $\langle
		T \rangle$ is a new result different from previous reports \cite{YaoKor,
			YaoKorLIDAR}. The difference comes from our extended consideration regarding $A$,
		$B$ and $d_r$ varying with $<T>$ and $<S>$.
	
	\bibitem{Jamieson}
	D.~Jamieson, J.~Tudhope, R.~Morris, and G.~Cartwright, \enquote{Physical
		properties of sea water solutions: heat capacity,}
	{\protect\JournalTitle{Desalination}} \textbf{7}, 23 -- 30 (1969).
	
	\bibitem{Nayar}
	K.~G. Nayar, M.~H. Sharqawy, L.~D. Banchik, and J.~H.~Lienghard,
	\enquote{Thermophysical properties of seawater: A review and new correlations
		that include pressure dependence,} {\protect\JournalTitle{Desalination}}
	\textbf{390}, 1 -- 24 (2016).
	
	\bibitem{ThC}
	D.~Jamieson and J.~Tudhope, \enquote{Physical properties of sea water
		solutions: thermal conductivity,} {\protect\JournalTitle{Desalination}}
	\textbf{8}, 393 -- 401 (1970).
	
	\bibitem{DV1}
	M.~H. Sharqawy, J.~H.~L. V, and S.~M. Zubair, \enquote{Thermophysical
		properties of seawater: a review of existing correlations and data,}
	{\protect\JournalTitle{Desalination and Water Treatment}} \textbf{16},
	354--380 (2010).
	
	\bibitem{DV4}
	F.~J. Millero, \enquote{Seawater as a multicomponent electrolyte solution,}
	{\protect\JournalTitle{The sea}} \textbf{5}, 3--80 (1974).
	
	\bibitem{density2}
	J.~Isdale and R.~Morris, \enquote{Physical properties of sea water solutions:
		density,} {\protect\JournalTitle{Desalination}} \textbf{10}, 329 -- 339
	(1972).
	
	\bibitem{density3}
	F.~J. Millero and A.~Poisson, \enquote{International one-atmosphere equation of
		state of seawater,} {\protect\JournalTitle{Deep Sea Research Part A.
			Oceanographic Research Papers}} \textbf{28}, 625 -- 629 (1981).
	
\end{thebibliography}
\end{document}